\documentclass[a4]{article}
\usepackage[osf]{libertine}
\usepackage{microtype}
\usepackage{array,booktabs}
\usepackage{amsmath}
\usepackage{amsthm}
\usepackage{amssymb}
\usepackage{amsfonts}
\usepackage{float}
\usepackage{xcolor}
\usepackage[T1]{fontenc}
\usepackage{longtable}
\usepackage[normalem]{ulem}
\usepackage{algorithm}
\usepackage[noend]{algpseudocode}
\usepackage[utf8]{inputenc}
\usepackage{enumitem}
\usepackage{graphicx}
\usepackage{colortbl}
\usepackage{tikz}
\usetikzlibrary{arrows,matrix,fit,positioning}

\usepackage{listings}
\newtheorem{theorem}{Theorem}
\newtheorem{lemma}[theorem]{Lemma}

\newtheorem{definition}[theorem]{Definition}

\newtheorem{example}[theorem]{Example}
\newtheorem{remark}[theorem]{Remark}

\newtheorem{attack}[theorem]{Attack}

\setlist{nosep} 

\providecommand{\keywords}[1]{\textbf{\textit{Keywords---}} #1}

\usepackage{xspace}
\usepackage{numprint}

\usepackage[colorinlistoftodos]{todonotes}
\usepackage{marginnote}

\usepackage{url,amsmath}
\usepackage{booktabs}
\usepackage{multirow}

\DeclareMathOperator\GL{GL}

\DeclareMathOperator{\fmap}{\mathfrak{F}\xspace}
\DeclareMathOperator{\pmap}{\mathfrak{P}\xspace}

\newcommand{\F}{\ensuremath{\mathbb{F}}}

\newcommand{\N}{\ensuremath{\mathbb{N}}}

\definecolor{mygreend}{HTML}{2ca92c}
\definecolor{myredd}{HTML}{c31313}

\usepackage{caption}
\definecolor{codegreen}{rgb}{0,0.6,0}
\definecolor{codegray}{rgb}{0.5,0.5,0.5}
\definecolor{codepurple}{rgb}{0.58,0,0.82}
\definecolor{backcolour}{rgb}{0.99,0.99,0.99}

\setlist[enumerate]{topsep=1ex, itemsep=1ex}


\begin{document}
\title{Breaking the Hidden Irreducible Polynomials Scheme}
%
%
\author{Christian Eder\\Department of Mathematics\\TU Kaiserslautern,
    Germany\\ederc@mathematik.uni-kl.de}
%
%
%
\maketitle              
\begin{abstract}
In~\cite{cryptoeprint:2019:1174} G\'omez describes a new public key
cryptography scheme based on ideas from multivariate public key cryptography
using hidden irreducible polynomials. We show that the scheme's design has a
flaw which lets an attacker recover the private key directly from the public
key.\\[0.3em]
\end{abstract}

\section{Introduction}
For several decades public-key cryptography schemes whose security based on
the hardness of solving multivariate polynomial systems over finite fields.
One of the first such schemes was in $1988$ $C^*$ by Matsumoto and
Imai~\cite{matsumoto-imai}, which was broken by Patarin in
$1995$~\cite{patarin-1995}. From this point onwards many multivariate
schemes, mostly signature schemes, evolved, for example, HFE~\cite{hfe-1996},
FLASH~\cite{flash-2001}, UOV~\cite{uov-1999}. These and other systems come
in many different variations of these systems. Especially multivariate
signature schemes are stand the test of time, some of them also part of
the ongoing post-quantum standardization process by the National Institute
of Standards and Technology (NIST).

Still, designing multivariate
encryption schemes is a harder task since most of the proposed systems
have been successully analyzed and broken.
In~\cite{cryptoeprint:2019:1174} G\'omez describes a new public key
cryptography scheme based on ideas from multivariate public key cryptography
using hidden irreducible polynomials. The fundamental idea behind the system
is polynomial multiplication and factorization.

\subsection{Our Contribution}
We show that the design of the Hidden Irreducible Polynomials scheme itself
reveals the private key which leads to a full break. We state two possible attacks.

\subsection{Organization}
The paper is organized as follows:
In Section~\ref{sec:scheme} we shortly review the Hidden Irreducible
Polynomials scheme. In Section~\ref{sec:break} we show how the private
key is extracted from the construction of the scheme. We then give two
possible attacks: One based on linear algebra, the other one directly
reading off the private key from the public one. In Section~\ref{sec:conclusion}
we conclude the full break of the scheme.

\section{Description of the scheme}
\label{sec:scheme}
We start with a short review of the construction of the Hidden Irreducible Polynomials
scheme:
\begin{definition} \
\label{def:scheme}
\begin{enumerate}
\item Let $p$ be a prime number, for a given $m \in \N$ we set $q :=
p^m$.\footnote{In~\cite{cryptoeprint:2019:1174} the author uses $n$ instead of
    $m$. Since $m$ denotes also a different property
        in~\cite{cryptoeprint:2019:1174}, we made the distinction by
        using different letters.}  We consider
the field extension $\F_{q^n} \cong \F_{q}[x] / h(x) =: K$ for some irreducible polynomial
$h \in \F_q[x]$ of degree $\deg(h) = n$.

\item We fix two polynomials
\begin{center}
$
\begin{array}{rcccccccc}
f(x) &:=& y_1 &+& y_2 x &+& \cdots &+& y_{k+1} x^k,\\
g(x) &:=& y_{k+2} &+& y_{k+3} x &+& \cdots &+& y_{2(k+1)} x^k.
\end{array}
$
\end{center}
in $K$ of degree $\deg(f) = \deg(g) = k$ for some prime number $k \in \N$
such that $2k < n-1$ and $y_i \in \F_q$ for all $1 \leq j \leq 2(k+1)$.\footnote{In~\cite{cryptoeprint:2019:1174}
the authors uses $p(x)$ and $q(x)$ instead of $f(x)$ and $g(x)$. Again, $p(x)$
    and $q(x)$ denote in~\cite{cryptoeprint:2019:1174} other polynomials.}
\label{def:polys}
\item The \textbf{private map} $\fmap$ is given by
\begin{center}
$
\begin{array}{rccc}
\fmap:& \F_{q^n} \times \F_{q^n} &\rightarrow &\F_{q^n},\\
        & \left(f(x),g(x)\right) & \mapsto & f(x) \cdot g(x).
\end{array}
$
\end{center}
\label{private-map}
\item For $u(x) = c_0 + c_1 x + \cdots c_{n-1} x^{n-1} \in K$
we define the one-to-one map
\begin{center}
$
\begin{array}{rccc}
\varphi:& K &\rightarrow &\F_q^n,\\
        & u(x) & \mapsto & (c_0,\ldots,c_{n-1}).
\end{array}
$
\end{center}
\item $T \in \GL\left(\F_q, (2k+1) \times (2k+1)\right)$ denotes the transformation matrix.
\item The \textbf{public map} $\pmap$ is given by:
\begin{center}
$
\begin{array}{rccc}
\pmap:& \F_{q}^{2(k+1)} &\rightarrow &\F_{q^n},\\
      & \left(t_1,\ldots, t_{2(k+1)}\right) & \mapsto & \sum_{i=0}^{2k}
      p_i\left(t_1,\ldots,t_{2(k+1)}\right) x^i.
\end{array}
$
\end{center}
where the $p_i \in \F_q[y_1,y_2,\ldots, y_{2(k+1)}]$ are constructed via
\begin{center}
$
\begin{array}{c}
\left(p_1\left(y_1,\ldots,y_{2(k+1)}\right), \ldots,
                p_{2k+1}\left(y_1,\ldots,y_{2(k+1)}\right),0,\ldots,0\right)\\
        := \\
\varphi^{-1} \circ T \circ \varphi \circ \fmap\left(f(x), g(x)\right)
\end{array}
$
\end{center}
with $n-2k-1$ zeroes at the end.
\label{def:pmap}
\end{enumerate}
\end{definition}

Applying the scheme for encryption and decryption one needs to implement
the following steps. Here we assume that Alice generates the private and the public map and distributes the public map. Bob now uses the public map to encrypt a message to Alice, which she then decrypts using the private map.
\begin{definition} \
\label{def:enc-dec}
\begin{enumerate}
\item With the above definitions Alice would create the private map
$\fmap$, construct a random transformation matrix $T$ and generate from these a corresponding
public map $\pmap$.
\item Bob is able to use Alice's public map $\pmap$: He
constructs two irreducible polynomials $p, q \in K$ both of degree $k$.
Bob wants to share $p$ and $q$ with Alice secretly.
\item Using the map $\varphi$, Bob can map the coefficients of $p$ and $q$ to
coefficient vectors in $\F_q^{k+1} \subset \F_q^n$. Concatenating $\varphi(p)$
and $\varphi(q)$ we receive a coefficient vector $v := \varphi(p) || \varphi(q) \in
\F_q^{2(k+1)}$. In other words, the information of both polynomials is 
encoded in one long vector of corresponding coefficients.
\item Bob uses $\pmap$, the system of quadratic multivariate polynomial
equations. Each polynomial $p_i$ in $\pmap$ takes $2(k+1)$ variables, so Bob
applies $v$ to all the $p_i$ and gets an element $w:=\pmap(v) \in \F_q^{2k+1}$.
Bob further applies $\varphi^{-1}$ to $w$ to receive the encrypted message $z
\in K$, a univariate polynomial of degree $\deg(z) = 2k$.
\item\label{dec} Bob sends $z$ to Alice. Alice first uses $\varphi$ to get the
coefficient vector of $z$. Then she can apply the inverse of the privately known
transformation matrix $T$. Finally, applying $\varphi^{-1}$, she receives a
polynomial $r \in K$ of degree $\deg(r) = 2k$. This univariate polynomial can now
be factorized, and Alice receives Bob's input polynomials $p,q$.
\end{enumerate}
\end{definition}

\begin{remark}\
\begin{enumerate}
\item One would assume due to the idea
of the scheme, that $p$ and $q$ are multiplied to a polynomial $r(x) = p(x)
\cdot q(x)$ and then $v$ is the coefficient vector of $r$. This is done under
the hood, as $\fmap$ is nothing else but multiplying the input polynomials
which are encoded as one long coefficient vector.
\item Note that the factorization step also does not hold any private
information: If the factorization would not be efficient, Alice could not
recover Bob's $p$ and $q$. So anyone who gets $r$ also gets $p$ and $q$.
\end{enumerate}
\label{rem:mult}
\end{remark}

\section{Breaking the scheme}
\label{sec:break}
In the last section we gave a review on how the hidden irreducible polynomials
scheme is constructed, how encryption and decryption works. There are two main
observations:

\begin{remark} \
\begin{enumerate}
\item $f,g$, the ingredients to construct the private map are known and unique
once $k$ is fixed. The coefficients $y_i$ cannot be further specified but need
to be parameters in order to be used in the public map $\pmap$ as the variables of
the multivariate quadratic polynomials $p_i$. So $\fmap$ is known to anybody.
\item Looking again at Step~\ref{dec} in Definition~\ref{def:enc-dec} Alice only
uses $\varphi$ (publicly known) and $T$ to receive $r$. Thus the only secret
part of the scheme is $T$, an invertible matrix
\end{enumerate}
\end{remark}

This leads to the first possible attack.

\begin{attack}[Using linear algebra only] \
By definition it holds that
\[\pmap = \varphi^{-1} \circ T \circ \varphi \circ \fmap.\]
Thus we can get $T$ via linear algebra computing
\[\varphi \circ \pmap = T \circ \left(\varphi \circ \fmap\right).\]
Here, all data besides $T$ is publicly known.
\end{attack}

It turns out that we do not even need to relinearize the system via defining
new variables $y_{i,j} := y_i y_j$ and solve the system of linear equations,
we can do even better.

We have seen in Definition~\ref{def:scheme}.\ref{private-map} that $\fmap$ consists
of the product $r$ of the two arbitrary univariate polynomials $f$ and $g$, both of
degree $k$, so we get
\begin{center}
$
\begin{array}[]{rcl}
r(x) &=& y_{1,k+2} x^0\\
       &+& \left(y_{1,{k+3}} + y_{2,k+2}\right) x^1\\
       &+& \left(y_{1,k+4} + y_{2,k+3} + y_3{,k+2}\right) x^2\\
       &+& \vdots\\
       &+& \left(y_{k,2k+2} + y_{k+1, 2k+1}\right) x^{2k-1}\\
       &+& y_{k+1,2k+2} x^{2k}.
\end{array}
$
\end{center}

For the sake of an easier notation let us define the following notion.

\begin{definition}
Let $k \in \N$..
We define the \textbf{$m$th coefficient
sum} to be
\[Y_m := \sum_{(i,j) \in I_m} y_{i,j}\]
such that $I_m := \left\{(i,j) \mid 1 \leq i \leq k+1, k+2 \leq j \leq 2k+2, i+j =
m+k+2\right\}$ for $1 \leq m \leq 2k+1$.
\label{def:coeffs}
\end{definition}

With Definition~\ref{def:coeffs} we can represent $r(x)$ in a more natural way:
\begin{equation}
\label{eq:r}
r(x) = \sum_{i=0}^{2k} Y_{i+1} x^i.
\end{equation}

Even more, we can easily proof the following statement.
\begin{lemma}
\label{lemma:disjoint}
$I_\ell \cap I_m = \emptyset \iff \ell \neq m$.
\end{lemma}
\begin{proof}
Both directions follow directly by the structure of $f$ and $g$
(Definition~\ref{def:scheme}.\ref{def:polys}) and the definition of $r(x) = f(x)
\cdot g(x)$.\qed
\end{proof}

This new representation of the main data structures of the hidden irreducible
polynomials scheme leads to another attack.

\begin{attack}[Reading off $T$ from $\pmap$]
\label{attack:read}
Applying $\varphi$ to $\fmap$ we get with Equation~\ref{eq:r}
\[\varphi \circ \fmap = \begin{pmatrix} Y_1\\Y_2\\ \vdots \\ Y_{2k} \\ Y_{2k+1}
\end{pmatrix}
\in \F_q^{2k+1} \subset \F_q^n.\]

By Definition~\ref{def:scheme}(\ref{def:pmap}) we have that
$\varphi \circ \pmap = T \circ \varphi \circ \fmap$. Thus rewriting $\varphi
\circ \pmap$ using the notation of coefficient sums
(Definition~\ref{def:coeffs})
we get
\[\varphi \circ \pmap =
\begin{pmatrix}
\sum_{\ell = 1}^{2k+1} t_{1,\ell} Y_\ell\\
\vdots\\
\sum_{\ell = 1}^{2k+1} t_{2k+1,\ell} Y_\ell
\end{pmatrix}.
\]
Since by Lemma~\ref{lemma:disjoint} all $I_\ell$ are disjoint, given $\pmap$, none of the
appearing coefficients in front of the $Y_\ell$ are interfered, but exactly the
matrix entries $t_{i,j}$. Thus, we can directly read off $T$ from $\pmap$.
\end{attack}

Let us recall the example given in Section~$6$ in~\cite{cryptoeprint:2019:1174}
to show how Attack~\ref{attack:read} works:

\begin{example}
In the given example we have $q=2$ and $k=7$. $T$ is thus given as the $15 \times 15$
$\F_2$-matrix
\begin{center}
\begin{tikzpicture}[scale=0.8]

\matrix[left delimiter={(}, right delimiter={)}, matrix of nodes] (mat) 
{
$0$ & $1$ & $0$ & $1$ & $0$ & $0$ & $0$ & $1$ & $0$ & $1$ & $1$ & $1$ & $1$ & $1$ & $1$\\
$1$ & $1$ & $1$ & $1$ & $1$ & $0$ & $0$ & $1$ & $1$ & $1$ & $1$ & $0$ & $0$ & $1$ & $0$\\
$1$ & $1$ & $1$ & $0$ & $1$ & $1$ & $1$ & $1$ & $1$ & $0$ & $1$ & $1$ & $1$ & $1$ & $1$\\
$1$ & $1$ & $0$ & $1$ & $0$ & $0$ & $1$ & $0$ & $1$ & $1$ & $1$ & $0$ & $0$ & $0$ & $1$\\
$1$ & $0$ & $0$ & $1$ & $1$ & $1$ & $0$ & $0$ & $1$ & $1$ & $0$ & $0$ & $1$ & $1$ & $1$\\
$1$ & $1$ & $0$ & $1$ & $0$ & $1$ & $0$ & $0$ & $1$ & $0$ & $0$ & $1$ & $1$ & $0$ & $0$\\
$1$ & $0$ & $0$ & $0$ & $1$ & $0$ & $1$ & $0$ & $1$ & $1$ & $0$ & $0$ & $0$ & $0$ & $1$\\
$1$ & $0$ & $0$ & $0$ & $0$ & $0$ & $1$ & $1$ & $1$ & $1$ & $0$ & $0$ & $0$ & $1$ & $1$\\
$0$ & $1$ & $0$ & $1$ & $1$ & $1$ & $1$ & $1$ & $0$ & $1$ & $1$ & $0$ & $1$ & $0$ & $1$\\
$0$ & $0$ & $0$ & $0$ & $1$ & $0$ & $0$ & $0$ & $1$ & $1$ & $1$ & $1$ & $1$ & $0$ & $0$\\
$1$ & $0$ & $1$ & $0$ & $1$ & $1$ & $0$ & $0$ & $0$ & $1$ & $1$ & $0$ & $0$ & $0$ & $1$\\
$1$ & $1$ & $0$ & $1$ & $1$ & $0$ & $0$ & $1$ & $1$ & $0$ & $1$ & $1$ & $1$ & $1$ & $0$\\
$1$ & $0$ & $0$ & $1$ & $1$ & $1$ & $0$ & $0$ & $1$ & $1$ & $1$ & $1$ & $0$ & $0$ & $1$\\
$1$ & $0$ & $0$ & $1$ & $0$ & $1$ & $0$ & $1$ & $0$ & $1$ & $1$ & $0$ & $0$ & $1$ & $1$\\
$1$ & $1$ & $0$ & $1$ & $1$ & $1$ & $1$ & $0$ & $0$ & $1$ & $0$ & $1$ & $0$ & $1$ & $0$\\
};
\node[left of=mat-7-1] at ([xshift=-0.5em, yshift=-0.5em]mat-7-1.west) {$T= $};
\end{tikzpicture}
\end{center}
Looking at $\pmap$ we get the following system of $15$ multivariate quadratic equations in the variables
$y_1,\ldots,y_{16}$:
\small{
\begin{center}
$
\begin{array}{l}
y_{2}y_{9}+y_{4}y_{9}+y_{8}y_{9}+y_{1}y_{10}+y_{3}y_{10}+y_{7}y_{10}+y_{2}y_{11}+y_{6}y_{11}+y_{8}y_{11}\\
+y_{1}y_{12}+y_{5}y_{12}+y_{7}y_{12}+y_{8}y_{12}+y_{4}y_{13}+y_{6}y_{13}+y_{7}y_{13}+y_{8}y_{13}+y_{3}y_{14}\\
+y_{5}y_{14}+y_{6}y_{14}+y_{7}y_{14}+y_{8}y_{14}+y_{2}y_{15}+y_{4}y_{15}+y_{5}y_{15}+y_{6}y_{15}+y_{7}y_{15}\\
+y_{8}y_{15}+y_{1}y_{16}+y_{3}y_{16}+y_{4}y_{16}+y_{5}y_{16}+y_{6}y_{16}+y_{7}y_{16}+y_{8}y_{16},\\[0.2em]
y_{1}y_{9}+y_{2}y_{9}+y_{3}y_{9}+y_{4}y_{9}+y_{5}y_{9}+y_{8}y_{9}+y_{1}y_{10}+y_{2}y_{10}+y_{3}y_{10}\\
+y_{4}y_{10}+y_{7}y_{10}+y_{8}y_{10}+y_{1}y_{11}+y_{2}y_{11}+y_{3}y_{11}+y_{6}y_{11}+y_{7}y_{11}+y_{8}y_{11}\\
+y_{1}y_{12}+y_{2}y_{12}+y_{5}y_{12}+y_{6}y_{12}+y_{7}y_{12}+y_{8}y_{12}+y_{1}y_{13}+y_{4}y_{13}+y_{5}y_{13}\\
+y_{6}y_{13}+y_{7}y_{13}+y_{3}y_{14}+y_{4}y_{14}+y_{5}y_{14}+y_{6}y_{14}+y_{2}y_{15}+y_{3}y_{15}+y_{4}y_{15}\\
+y_{5}y_{15}+y_{8}y_{15}+y_{1}y_{16}+y_{2}y_{16}+y_{3}y_{16}+y_{4}y_{16}+y_{7}y_{16},\\[0.2em]
\vdots\\[0.2em]
y_{1}y_{9}+y_{4}y_{9}+y_{6}y_{9}+y_{8}y_{9}+y_{3}y_{10}+y_{5}y_{10}+y_{7}y_{10}+y_{2}y_{11}+y_{4}y_{11}\\
+y_{6}y_{11}+y_{8}y_{11}+y_{1}y_{12}+y_{3}y_{12}+y_{5}y_{12}+y_{7}y_{12}+y_{8}y_{12}+y_{2}y_{13}+y_{4}y_{13}\\
+y_{6}y_{13}+y_{7}y_{13}+y_{1}y_{14}+y_{3}y_{14}+y_{5}y_{14}+y_{6}y_{14}+y_{2}y_{15}+y_{4}y_{15}+y_{5}y_{15}\\
+y_{8}y_{15}+y_{1}y_{16}+y_{3}y_{16}+y_{4}y_{16}+y_{7}y_{16}+y_{8}y_{16},\\[0.2em]
y_{1}y_{9}+y_{2}y_{9}+y_{4}y_{9}+y_{5}y_{9}+y_{6}y_{9}+y_{7}y_{9}+y_{1}y_{10}+y_{3}y_{10}+y_{4}y_{10}\\
+y_{5}y_{10}+y_{6}y_{10}+y_{2}y_{11}+y_{3}y_{11}+y_{4}y_{11}+y_{5}y_{11}+y_{8}y_{11}+y_{1}y_{12}+y_{2}y_{12}\\
+y_{3}y_{12}+y_{4}y_{12}+y_{7}y_{12}+y_{1}y_{13}+y_{2}y_{13}+y_{3}y_{13}+y_{6}y_{13}+y_{8}y_{13}+y_{1}y_{14}\\
+y_{2}y_{14}+y_{5}y_{14}+y_{7}y_{14}+y_{1}y_{15}+y_{4}y_{15}+y_{6}y_{15}+y_{8}y_{15}+y_{3}y_{16}+y_{5}y_{16}+y_{7}y_{16}.
\end{array}
$
\end{center}
}
\ \\
Simply applying Lemma~\ref{def:coeffs} the system gets way easier:
\begin{center}
$
\begin{array}{l}

Y_{1}+Y_{3}+Y_{7}+Y_{9}+Y_{10}+Y_{11}+Y_{12}+Y_{13}+Y_{14},\\
Y_{0}+Y_{1}+Y_{2}+Y_{3}+Y_{4}+Y_{7}+Y_{8}+Y_{9}+Y_{10}+Y_{13},\\
\vdots\\
Y_{0}+Y_{3}+Y_{5}+Y_{7}+Y_{9}+Y_{10}+Y_{13}+Y_{14},\\
Y_{0}+Y_{1}+Y_{3}+Y_{4}+Y_{5}+Y_{6}+Y_{9}+Y_{11}+Y_{13}.
\end{array}
$
\end{center}

Writing down, for example, $p_{14}$ (second to last one) in a dense representation we get:
\[
1 \cdot Y_{0} + 0 \cdot Y_{1} + 0 \cdot Y_{2} + 1 \cdot Y_{3} + 0 \cdot Y_{4} + 1 \cdot Y_{5} + 0 \cdot Y_{6}
+1 \cdot Y_{7}+ 0 \cdot Y_{8} + 1 \cdot Y_{9} + 1 \cdot Y_{10} + 0 \cdot Y_{11} + 0 \cdot Y_{12}
+ 1 \cdot Y_{13}+ 1 \cdot Y_{14}.
\]
Reading off the corresponding coefficient vector we get exactly the $14$th row of $T$:
\begin{center}
\begin{tikzpicture}[]

\matrix[left delimiter={[}, right delimiter={]}, matrix of nodes] (mat) 
{
$1$ & $0$ & $0$ & $1$ & $0$ & $1$ & $0$ & $1$ & $0$ & $1$ & $1$ & $0$ & $0$ & $1$ & $1$\\
};
\end{tikzpicture}.
\end{center}
\end{example}

\section{Conclusion}
\label{sec:conclusion}
In this paper we have shown a full break of the Hidden Irreducible Polynomials
scheme introduced by G\'omez in~\cite{cryptoeprint:2019:1174}. We have shown
that the private key is publicly known by the design of the system. Moreover, we
have shown that due to the construction of the private map, namely univariate
polynomial multiplication, one can even easily read off the transformation
matrix for the system of multivariate quadratic polynomial equations such that
not even linear algebra is needed for attacking the scheme.


\begin{thebibliography}{1}

\bibitem{cryptoeprint:2019:1174}
Borja Gómez.
\newblock Hidden irreducible polynomials : A cryptosystem based on multivariate
  public key cryptography.
\newblock Cryptology ePrint Archive, Report 2019/1174, 2019.

\bibitem{uov-1999}
Aviad Kipnis, Jacques Patarin, and Louis Goubin.
\newblock Unbalanced oil and vinegar signature schemes.
\newblock In {\em Proceedings of the 17th International Conference on Theory
  and Application of Cryptographic Techniques}, EUROCRYPT'99, pages 206--222,
  Berlin, Heidelberg, 1999. Springer-Verlag.

\bibitem{matsumoto-imai}
T.~Matsumoto and H.~Imai.
\newblock Public quadratic polynomial-tuples for efficient
  signature-verification and message-encryption.
\newblock In {\em Lecture Notes in Computer Science on Advances in
  Cryptology-EUROCRYPT'88}, pages 419--453, New York, NY, USA, 1988.
  Springer-Verlag New York, Inc.

\bibitem{patarin-1995}
Jacques Patarin.
\newblock Cryptoanalysis of the matsumoto and imai public key scheme of
  eurocrypt'88.
\newblock In {\em Proceedings of the 15th Annual International Cryptology
  Conference on Advances in Cryptology}, CRYPTO '95, pages 248--261, Berlin,
  Heidelberg, 1995. Springer-Verlag.

\bibitem{hfe-1996}
Jacques Patarin.
\newblock Hidden fields equations (hfe) and isomorphisms of polynomials (ip):
  Two new families of asymmetric algorithms.
\newblock In {\em Proceedings of the 15th Annual International Conference on
  Theory and Application of Cryptographic Techniques}, EUROCRYPT'96, pages
  33--48, Berlin, Heidelberg, 1996. Springer-Verlag.

\bibitem{flash-2001}
Jacques Patarin, Nicolas Courtois, and Louis Goubin.
\newblock Flash, a fast multivariate signature algorithm.
\newblock In {\em Proceedings of the 2001 Conference on Topics in Cryptology:
  The Cryptographer's Track at RSA}, CT-RSA 2001, pages 298--307, Berlin,
  Heidelberg, 2001. Springer-Verlag.

\end{thebibliography}

\end{document}